\documentclass[12pt,preprint]{aastex}

\begin{document}


\title{The Hall instability of weakly ionized, radially stratified,
rotating disks}


\author{Edward Liverts and Michael Mond}
\affil{Department of Mechanical Engineering,  Ben-Gurion
University of the Negev, \\ P.O. Box 653, Beer-Sheva 84105,
Israel}

\email{eliverts@bgu.ac.il}

\and

\author{Arthur D. Chernin}
\affil{Sternberg Astronomical Institute, Moscow
 State University, \\ University Prospect 13,
 Moscow 119899, Russia}




\begin{abstract}
Cool weakly ionized gaseous rotating disk, are considered by many
models as the origin of the evolution of protoplanetary clouds.
Instabilities against perturbations in such disks play an
important role in the theory of the formation of stars and
planets. Thus, a hierarchy of successive fragmentations into
smaller and smaller pieces as a part of the Kant-Laplace theory of
formation of the planetary system remains valid also for
contemporary cosmogony. Traditionally, axisymmetric
magnetohydrodynamic (MHD), and recently Hall-MHD instabilities
have been thoroughly studied as providers of an efficient
mechanism for radial transfer of angular momentum, and of density
radial stratification. In the current work, the Hall instability
against nonaxisymmetric  perturbations in compressible rotating
fluid in external magnetic field is proposed as a viable mechanism
for the azimuthal fragmentation of the protoplanetary disk and
thus perhaps initiating the road to planet formation. The Hall
instability is excited due to the combined effect of the radial
stratification of the disk and the Hall electric field, and its
groth rate is of the order of the rotation period. Such family
of instabilities are introduced here for the first time in
astrophysical context.
\end{abstract}


\keywords{protoplanetary disks --- instabilities --- MHD}



\section{Introduction}\label{sec:intro}
Linear mode analysis provides a useful tool for gaining important
insight into the relevant physical processes that determine the
stability of rotating fluid configurations. The importance of
magnetic fields in rotating disks has been demonstrated by the
rediscovery of the magneto rotational instability (MRI) in which
hydrodynamically stable flows with an angular velocity that is
decreasing outwards are highly unstable when threaded by a weak
magnetic field \cite{bh}. That investigation has been carried out
in the magnetohydrodynamic (MHD) limit and invoked a number of
approximations appropriate to the study of the evolution of
long-wavelength perturbations in the weak magnetic field limit.
The inclusion of the Hall electric field (Hall MHD) is a
relatively recent development
\cite{W99,BT,sa,sb,SW03,Desch,ur,rk}. The Hall electric field
plays an important role in the disk's dynamics when the coupling
between the electrons and neutral components of the fluid is low.
In such cases, the inertial length of the ions is longer than the
characteristic perturbation's length scale and consequently the
motions of the ions and electrons are decoupled. Indeed, it has
been shown that the Hall electric field has a profound effect on
the structure and grouth rate of unstable modes like the MRI's.
Furthermore, as will be shown below, the Hall term gives rise to
new branches of unstable modes. In particular, it will be shown
that the Hall term, in the presence of radial stratification,
excites non axisymmetric instabilities.

As an astrophysical interest, we mention magnetically supported
cool molecular clouds and its dynamics. In the disk of typical
spiral galaxy, the magnetic field strength is usually estimated to
be of several to more than $10 \mu G$  while in some regions of
spiral galaxies the magnetic field strength may be higher than
several tens of microgauss \citep[see][]{sf,Beck}. By that
estimations and others, it is almost certain that MHD density
waves should also play  an important role in the dynamic and
evolution of structures  within a magnetized gas disk \cite{Fan}.
As is well known in classical nebular hypothesis by Kant-Laplace
the condensation in protoplanetary rotating disk plays an
important part in forming stars and planets. That part of the
Kant-Laplace theory remains valid also for a contemporary
cosmogony. The "standard" theory of the multistage accretionary
formation of planets, or the so-called core accretion mechanism
\cite{Safronov,Pollack} remained the most popular until recently,
when it was criticized by \cite{Boss02,Boss03} and others. The
main problem of the latter is the timescale, which is longer than
estimates of the lifetime of many planet-forming disks
\cite{Taylor,Feigelson}. In any case all theories rely on
instabilities as a mechanism to transform a relatively uniform
rotating gaseous disk into a planetary system. That is, at an
early stage, the protosolar nebula are formed by fragments that
separated from a molecular cloud. Planetary formation is thought
to start with inelastically colliding gaseous and dust particles
settling to the central plane of the rotating nebula to form a
thin layer around the plane. During the early evolution of the
disk it is believed that the dust particles coagulate also into
comets-planetesimals. On attaining a certain critical thickness
(and, correspondingly, low temperature) small in comparison with
the size of the disk, as a result of a local gravitational
collapse the nebula disintegrats into the central body and a
number of separate protoplanets. Instabilities arise as the
thickness of the disk is reduced \cite{Gurevich,Safronov}. If a
rotating gaseous disk has a large vertical thickness owing to a
high internal temperature, then it is stabilzed against
gravitational instabilities by thermal motion \cite{Gurevich}. In
\cite{Boss04} it is demonstrated that convective cooling is able
to cool the disk midplane at the desired rate to produce clumps in
marginally unstable disks. The physical phenomena treated in the
current paper occur during the stage of evolution of the
protoplanetary cloud when the dust and gas in the disk start to
condense into planetesimals and a star with current luminosity
emerges at the center of the nebula.

The rest of paper is organized as follows. In Sec.2.1 we present
basic equations and state our assumptions. In Sec.2.2 we present
the dispersion relation to be solved. In Sec.2.3 we pay particular
attention to the conditions of the existing of complex-conjugate
roots of the dispersion relation. We present our conclusions and
discussion in Sec.3.

\section{Hall MHD equations and the dispersion relation}
\subsection{Basic equations}
We consider a thin rotating gaseous disk with angular velocities
$\Omega(r)$, where $G$ is the gravitational constant and $M$ is
the mass of the central body, and $r$ is the distance from the
center of the rotating disk. The thickness of the disk can be
estimated by $c_s/\Omega(r)$ where $c_s$ is the sound speed. The
disk is made of partially ionized plasma where ions are well
coupled to the neutrals while the electrons are not. However,
charge neutrality, $n_e=n_i$, is assumed to be valid. The disk is
immersed in a magnetic field directed along the rotation axis
(defined as the $z$-axis in our frame of reference). Following
\cite{Braginskii}, the equations that govern the evolution of the
two fluid system, namely the heavy particles (ions and atoms) and
the electrons, are:

the momentum equation
\begin{equation}
nm_i{{d\vec{v}}\over {dt}}=-\vec{\nabla
p}+\frac{1}{c}\vec{j}\times\vec{B} -nm_i\frac{GM}{r^3}\vec{r},
\label{Moment}
\end{equation}
here $n$ is the number density of the heavy particles (ions and
atoms), and the generalized Ohm's law
\begin{equation}
m_e\frac{d\vec{u}}{d t}=e\vec{E}+\frac{e}{c}\vec{v}\times\vec{B}
-\frac{1}{n_ec}\vec{j}\times\vec{B}+T_e\frac{\vec{\nabla
n_e}}{n_e}-\frac{e\vec{j}}{\sigma_R}, \label{Om'sLaw}
\end{equation}
where $\sigma_R$ is the electrical conductivity.
The generalized Ohm's law as given in Eq.(\ref{Om'sLaw}) differs
from the corresponding equation of MHD theory by the term on the
left-hand side that describes the effect of electron inertia, by
the third term on the right-hand side, which is the
Hall effect, by the fourth term, which describes the effect of
the electrons' pressure and by the last term that represents the drag
force acting on the electrons. In addition, it is convenient to write
the induction equation by substituting the electric field from
Eq.(\ref{Om'sLaw}) into Faraday's equation which than becomes
\begin{equation}
{\partial\vec{B}\over \partial t}=\vec{\nabla}\times(\vec{v}\times
\vec{B})-\vec{\nabla}\times({{\vec{j}\times\vec{B}}\over{en_e}})
-\frac{m_ec}{e}\vec{\nabla}\times\frac{d\vec{u}}{d t}+\eta\Delta
\vec{B},
\label{InductionLaw}
\end{equation}
where $\eta=c^2/4\pi\sigma_R$ is magnetic diffusivity.
The relative importance of the various terms in
Eq.(\ref{InductionLaw}) may be investigated by considering
appropriate length and time scales. Thus, consider, for example,
the second term on the right-hand side of Eq.(\ref{InductionLaw})
(the Hall term). If we assume that $\nabla \approx 1/L,\quad
j\approx cB/4\pi L$ (displacement current is neglected), where $L$
is a typical length-scale of the density inhomogeneity, then that
term will be of the same order as the convective electric field if
$\ell_i\approx\sqrt{\zeta} L$ where
$$
\ell_i=\frac{c}{\omega_{pi}}=\frac{c\sqrt{m_i}}{\sqrt{4\pi e^2
n_i}}.
$$
In other words, in order for the Hall term to be important the
length-scale $L$ should be of the same order of magnitude or less
than the ions inertial length divided by $\sqrt{\zeta}$ where
$\zeta$ is ionization degree, i.e.,
\begin{equation}
L\leq\ell_i/\sqrt{\zeta}. \label{cond}
\end{equation}
Similarly, the ratio of the electron inertia term [the last term
on right-hand side of Eq.(\ref{InductionLaw})] to the convective
term is given by:
$$
\frac{\omega}{\Omega_e}\frac{c}{4\pi e n_e L}\approx
\frac{1}{\sqrt{4\pi n m_i}}
$$
Thus, the electron's inertial term is important for
inhomogeneities which are characterized by $L\approx
c/\omega_{pe}=\ell_e$. As far as the time-scale is concerned, it
is seen from the relationship $v_A/\ell_i =\sqrt{\zeta}\Omega_i$
($v_A=B/\sqrt{4 \pi n m_i}$ is the Alfv\'{e}n velocity) that for
the Hall term to be important the frequency should be
$\omega>\sqrt{\zeta}\Omega_i$ while the electron inertia should be
retained if $\omega>\Omega_e$. Note that in case of
$\omega>>\sqrt{\zeta}\Omega_i$ the electrons drift in the wave's
electric field, while the ions are immobile. Thus, if
$\sqrt{\zeta}\Omega_i << \omega << \Omega_e$ the second term on
the right-hand side of Eq.(\ref{InductionLaw}), namely the Hall
term, is the leading term. This is the reason to term such
approximation HMHD and waves in that regime Hall waves. It is seen
from Eq.(\ref{InductionLaw}) that in such case
$$
\omega\approx\frac{c^2\Omega_i}{\omega_{pi}^2L^2}=
\frac{v_{Hd}}{L}
$$
where $ v_{Hd}=v_A^2/\zeta\Omega_i L$ is the phase velocity of the Hall
waves in the presence of density gradients whose scale length is
$L$. Such waves exist merely due to electron drift in the electric
field of the Hall waves. It should be noted from the
discussion above that the conditions for the Hall term to be
significant are more easily satisfied as the degree of ionization
is decreasing. Finally, it should be noted that for low enough densities
the Ohmic dissipation term in Eq.(3) (the last term on the right hand side)
is negligible in comparison with the Hall term \cite{BT}. Additional support
to that point of view is provided by \cite{Jin} who estimated the ratio of the
rotation period to Ohmic dissipation time to be of the order of $10^{-3}k^2H^2$.
As will be seen in the subsequent sections, the growth rates of the Hall
instability are of the order of the rotation period and the relevant wave
lengths satisfy $kH<1$. Hence, the effect of Ohmic dissipation will be neglected
from now on bearing in mind that it may nonetheless lower the growth rates of
the investigated instabilities.

\subsection{The linearized equations and the dispersion relation}

We consider a differentially rotating disk with angular velocity
$\Omega(r)$ where $r$ is the distance from the disk's center, and
$r$-dependent density $\rho(r)$. The disk is threaded by and axial
magnetic field $\vec{B}=B_0(r)\hat{z}$. The equations governing
the linear stability of the rotating disk may be derived from
Eqs.(1)-(3) by assuming that the perturbations of the steady
rotation are of the following form
\begin{equation}
f(r,\theta)=f(r)exp[i(m\theta-\omega t)] \label{lokal}
\end{equation}
where $f(r,\theta)$ stands for the perturbation of any of the
physical variables that describe the system. It should be noted
that strictly speaking expression (\ref{lokal}) may be used only
for rigid rotation, as differential rotation results in non
exponential perturbations. Nonetheless that fact accentuates the
nondependence of the Hall instability on the rotation shear, in
contrast to MHD instabilities like MRI's. A full description of
the temporal evolution of perturbations in differentially rotating
disk will be described in a forthcoming publication. The
amplitudes of the perturbed $\theta$-component of the velocity,
density and $z$-component of the magnetic field can be expressed
in terms of the amplitude of the radial component of the perturbed
velocity. This results in an ordinary differential equation for
$u_r(r)$ that should be solved with appropriate boundary
conditions thus yielding an eigenvalue problem. However for
purposes of demonstration we first consider the simplified case of
local approximation by assuming that $k_r \ll k$ where
$k_r\sim\frac{1}{u_r}\frac{\partial u_r}{\partial r}$ and
$k\sim\frac{m}{r}$. After linearization, Eqs.(1)-(3) assume the
following form:
\begin{equation}
i(\omega-m\Omega)u_r+2\Omega u_\theta\cong0 \label{u_r}
\end{equation}
\begin{equation}
(\omega-m\Omega)u_\theta+i\frac{\chi^2}{2\Omega}u_r=
\frac{m}{r}c^2_s\frac{\sigma}{\rho}+\frac{m}{r}V^2_A\frac{b_z}{B_0}
\label{u_theta}
\end{equation}
\begin{equation}
-\frac{1}{\zeta\rho\Omega_i}\frac{m}{r}\nabla_r\frac{B_0^2}{8\pi}
\frac{\sigma}{\rho}+\frac{m}{r}u_\theta-
[\omega-m(\Omega+\frac{{\ell}^2_i \Omega_i}{
Lr})]\frac{b_z}{B_0}=0 \label{b_z}
\end{equation}
\begin{equation}
(\omega-m\Omega)\frac{\sigma}{\rho}-\frac{m}{r}u_\theta=0
\label{div_u}
\end{equation}
where $\sigma$ is the perturbation of the density, and $\chi$ is
the epicyclic frequency given by $\chi = \sqrt {4\Omega ^2 +
2r\Omega d\Omega /dr}$. The system of equations (6)-(9) yields
following dispersion relation:
\begin{equation}
\tilde{\omega}^3-\tilde{\omega}^2\omega_{Hd}-\tilde{\omega}
[\chi^2+k^2(c^2_s+V^2_A)]
+\omega_{Hd}(\chi^2+k^2\frac{L}{\rho}\frac {\partial {\cal
P}}{\partial r})=0 \label{disp}
\end{equation}
where $\cal P$ is the total unperturbed pressure,
$\tilde{\omega}=\omega-m\Omega$, $k=m/r$, and
$\omega_{Hd}=m\frac{{\ell}^2_i \Omega_i}{ L r}$ is the Hall
drift frequency. It is a direct result of the assumed form of the
perturbation, i.e., Eq.(5) that the axial and radial components of the
perturbed magnetic field as well as the axial component of the
perturbed velocity are zero. In addition, the derivatives of
the equilibrium profiles have been neglected in deriving the linearized
equations above, except in the axial component of Faraday's law (Eq.(8))
where, due to the Hall term, they are of the same order of the rest
of the terms.
\subsection{The instability against non axisymmetric perturbations}
In the case of homogeneous density and magnetic field strength
($L\rightarrow\infty$) the two roots of Eq.(\ref{disp}) represent
two stable branches of density waves that originate due to both
the rotation of the disk as well as the external magnetic field.
However in the case of density or magnetic field inhomogeneity the
roots of Eq.(\ref{disp}) with real coefficients are real if, and
only if, the following conditions are satisfied:
\begin{equation}
D=\frac{\omega^6_{Hd}}{108}(27X^2+4X(1-9Y^2)-4Y^2+8Y^4-4Y^6)\leq 0
\end{equation}

where $$X=-\frac{k^2L\nabla_r{\cal P}}{\rho\omega^2_{Hd}}+
\frac{k^2(c^2_s+V^2_A)}{\omega^2_{Hd}}$$

and
$$Y^2=\frac{\chi^2}{\omega^2_{Hd}}+\frac{k^2(c^2_s+V^2_A)}{\omega^2_{Hd}}$$.
If the last condition is not fulfilled, Eq.(\ref{disp}) has two
complex-conjugate roots one of which signifies instability. The
conditions for that to happen are:
\begin{equation}
X\geq\frac{2}{27}(-1+9Y^2+\sqrt{1+9Y^2+27Y^4+27Y^6})
\end{equation}
or
\begin{equation}
X\leq\frac{2}{27}(-1+9Y^2-\sqrt{1+9Y^2+27Y^4+27Y^6})
\end{equation}

Further insight into the onset of the Hall instability may be
gained by denoting the total pressure radial derivative by
$(p+B_0^2/8\pi)/L_{\cal P}$ where $L_{\cal P}$ is the
inhomogeneity length of the total pressure. Next, the well known
relation $c_s=H\Omega$ is recalled, and finally normalizing
frequencies to $\Omega$, and wavelengths to $H$ the following
dispersion relation is obtained:
\begin{equation}
\tilde{\omega}^3-q\alpha\tilde{\omega}^2-\tilde{\omega}
[\hat{\chi}^2+q^2(1+\beta^{-2})] +q\alpha[\hat{\chi}^2+\xi
q^2(1+\beta^{-2}/2)]=0 \label{disp1}
\end{equation}
where
$$
\alpha=\frac{1}{\beta}\frac{\ell_i}{L\sqrt{\zeta}},
$$
$\beta=c_s/V_A$, $q=kH$, $\hat{\chi}=\chi/\Omega$, and
$\xi=L/L_{\cal P}$. It is first noted that for $\alpha\rightarrow
0$ the MHD regime is recovered and the roots of Eq. (\ref{disp1})
represent the stable combination of the fast magnetosonic waves
and the epicyclic oscillations. However, as $\alpha$ is increased
(which means that $L$ is decreased relative to the inertial length
of the ions) the system enters into the Hall MHD regime. In that
case, elementary analysis of Cardano's solution of cubic equations
reveals that the nature of the roots of the dispersion equation
(\ref{disp1}) hinges upon the value of $\mu$ which is given by:
\begin{equation}
\mu=\frac{\hat{\chi}^2+\xi q^2(1+\beta^{-2}/2)}{\hat{\chi}^2+
q^2(1+\beta^{-2})},
\end{equation}
and the roots of the following quadratic equation:
\begin{equation}
4\mu S^2+(1+18\mu -27\mu ^2)S+4=0, \label{quadratic}
\end{equation}
where
$$
S=\frac{q^2\alpha^2}{\hat{\chi}^2+ q^2(1+\beta^{-2})}.
$$
 Thus, Eq.
(\ref{disp1}) has two complex roots and hence the system is
unstable in the following two cases:
\begin{enumerate}
\item $\mu >1$, for $S_1<S<S_2$, where $S_1$ and $S_2$ are the
roots of Eq. (\ref{quadratic}). \\
In this case $\xi$ must be positive which means that the density
and the total pressure change radially in the same direction. It
is therefore obvious that regions of instability occur where the
total pressure changes more rapidly than the density ($\xi >1$).
This is indeed the case in polytropic disks for which
$L/L_p=\gamma>1$ where $L_p$ is the inhomogeneity length
associated with the pressure. Hence $\xi >\gamma$ and consequently
$\mu >1$, which means that the disk is unstable under the Hall
instability if $\beta$ is such that $S$ is between the two roots
of Eq. (\ref{quadratic}). Exact values of the growth rate,
obtained by the numerical solution of Eq. (\ref{disp1}), are
depicted in Fig. 1 for Keplerian rotation ($\hat{\chi}=1$), with
$\ell_i/L\sqrt {\zeta} =10$, and $\xi =5/3$, for various values of
$\beta$. It is found that the disk is unstable for $1<\beta <20$.
\item $\mu <0$, for $S>max(S_1,S_2)$.\\
In this case it is obvious that the gradients of the density and
the total pressure must have opposite signs (i.e. $\xi <0$). Such
situations may occur when radial inflow plays an important role in
the dynamics of the evolving disk, such as in young protoplanetary
clouds \cite{Hogerheijde}, or when radial rings of non monotonic
density profiles are formed due to gravitational instabilities
\cite{Mayer et al}. In these cases, in the limit $\alpha\gg 1$ one
of the solutions of Eq.(\ref{disp1}) is approximated by equating
the second and fourth terms in Eq.(\ref{disp1}) and is given by:
\begin{equation}
\omega=\pm i\gamma,
\end{equation}
where
\begin{equation}
\gamma=\Omega\sqrt{|\xi|q^2(1+\beta^{-2}/2)-\hat{\chi}^2}.
\end{equation}
It is clear that in this case, the rotation plays a stabilizing
role. Also, since $S\propto 1/\beta$, there is an upper bound on
$\beta$ for instability to occur but not a lower bound. Further
more, in the limit of small $\beta$ the growth rate grows without
bound as $\beta$ is decreased. Numerical solutions of Eq.
(\ref{disp1}) for Keplerian rotation, and for $\ell_i/L\sqrt
{\zeta} =10$ and $\xi =-1.5$ are depicted in Fig. 2. The
instability exists for $\beta <9$ and the growth rate is indeed a
growing function of $1/\beta$.
\end{enumerate}
The unstable branch of the dispersion relation (\ref{disp1}),
termed the Hall instability, has been investigated in detail in
\cite{LM} where it has been shown that it represents a
quasi-electrostatic slow mode in which the perturbation in the
ions' density and velocity play a crucial role while the perturbed
electrons' density and magnetic field are negligible. Thus, the
Hall instability provides a powerful mechanism for the azimuthal
breaking of radially stratified disks into small fragments of size
comparable to the disk's thickness and on time scale of the order
of the rotation period.

\section{Summary and discussion}

This paper examined the instability of a weakly ionized, thin
disks threaded by an external magnetic field within the Hall-MHD
model. The vertical stratification as well as the azimuthal
variations of the disks properties were ignored whereas radial
distribution was considered. In particular, in astrophysical
context such weakly ionized gaseous nebulae are relevant to the
protoplanetary disks. The conditions in protoplanetary disks were
discussed in ~\cite{Safronov,Gurevich}. At a certain stage of its
evolution the star nebula is believed to have a characteristic
disk size up to order of 30-100 AU and the total mass of the disk
is believed to be less than roughly 0.1 of the mass of the central
star. This yields an integrated column density of
$\Sigma\approx3\times10^2 g\cdot cm^{-2}$. Assuming that the mean
mass of the particles is $m_p=\Lambda m_H$, the number density is
given by $n=\Sigma/\Lambda m_H H(r)\simeq 2\times10^{14} cm^{-3}$
where $H(r)$ is the scale thickness of the disk. To estimate the
value of the factor $\Lambda$ it should be noted that within this
system, lighter elements such as hydrogen and helium were driven
out of the central regions by star wind and radiation pressure
during a highly active phase, leaving behind heavier elements like
Na, Al, and K and dust particles. Thus in the outer part of the
star nebula, ice and volatile gases were able to survive. As a
result, the inner planets are formed of minerals, while the outer
planets are more gaseous or icy. Concerning thickness estimation
one can use $c_s(r)/\Omega(r)$, which yields $H(r)\approx
0.002$AU. It should be noted also that due to very low temperature
of the protoplanetary disks the only sources of ionization are
non-thermal, e.g., cosmic rays, X-rays and the decay of
radioactive elements. Following \cite{sb} the ionization degree at
the midplane of gaseous disk is estimated as
$\zeta=n_e/n\approx10^{-12}$. So the disk material is a partially
ionized plasma where ions and charged small dust grains are well coupled
to the neutrals but electrons are not.
Following this assumption we can estimate that
$\ell_i/\sqrt{\zeta}$ is up to order of 1 AU. It should be noted that
such big values are obtained mostly due to low degree of ionization, however
one should keep in mind that existence of positivly
charged grain particles increases the inertial length (see definition
after Eq.(3)) and thus enhances the effect of the Hall term.
In Sec.2 it has been demonstrated that the Hall term is important if
$H(r)<L<\ell_i/\sqrt{\zeta}$. Thus, due to the very small values
of the ionization degree in protoplanetary disks the HMHD model
has to be employed when studying the stability of structures with
realistic radial density inhomogeneities. Indeed, radial
stratification with length scale $L>H$ may exist in the disk due
to such mechanisms as axisymmetric density waves that give rise to
alternating high and low density rings. On the other hand, $L$ is
bounded from below by $H$ due to thermal pressure \cite{Gurevich}.
Hence, as has been further shown in Sec.2, protoplanetary disks
with such radial density distributions are susceptible to strong
nonaxisymmetric instabilities whose growth rates are of the order
of the rotation period of the disk. Such instabilities result in
breaking of the density rings into fragments that may be
identified as planetesimals. Following widely adopted standpoint,
an accumulation of planetesimals may lead to the next stage of
evolution of protoplanetary disk, which is the coalesce of the
planetesimals into protoplanets. Such planetesimals may survive
the thermal pressure if their characteristic size, i.e., $1/k$, is
bigger the disk's thickness $H$ \cite{Gurevich}. On the other
hand, the linear analysis presented in Sec.2.2 is valid if $kL<1$.
Combining those two conditions, and taking into account condition
(\ref{cond}) results in the following limitations on $k$ :
\begin{equation}
\sqrt{\zeta}\frac{\Omega_i}{\Omega} < kH < 1
\label{ineq2}
\end{equation}
where value of $\beta$ has been taken as $1$ for simplicity.
Thus, for typical values in
protoplanetary disks, namely, $\zeta=10^{-12}$,
$\Omega_i=10^4\times B_0/G s^{-1}$, $\Omega\approx1.9\times 10^{-7} s^{-1}$ the
left hand side of the second inequality in (\ref{ineq2}) is of the order
of or smaller than unity. It is therefore again the small
ionization degree that enablles the onset of the Hall instability,
in the small magnetic field limit, and by thus providing a mechanism of
initiation of the standart scenario of planet formation.

\acknowledgments

A.C. thanks Prof. M.Mond and The Ben Gurion University for warm hospitality

This work has been supported by the Israel Science Foundation under
Contract No.265/00.

\clearpage



\begin{figure}
\epsscale{.80} \plotone{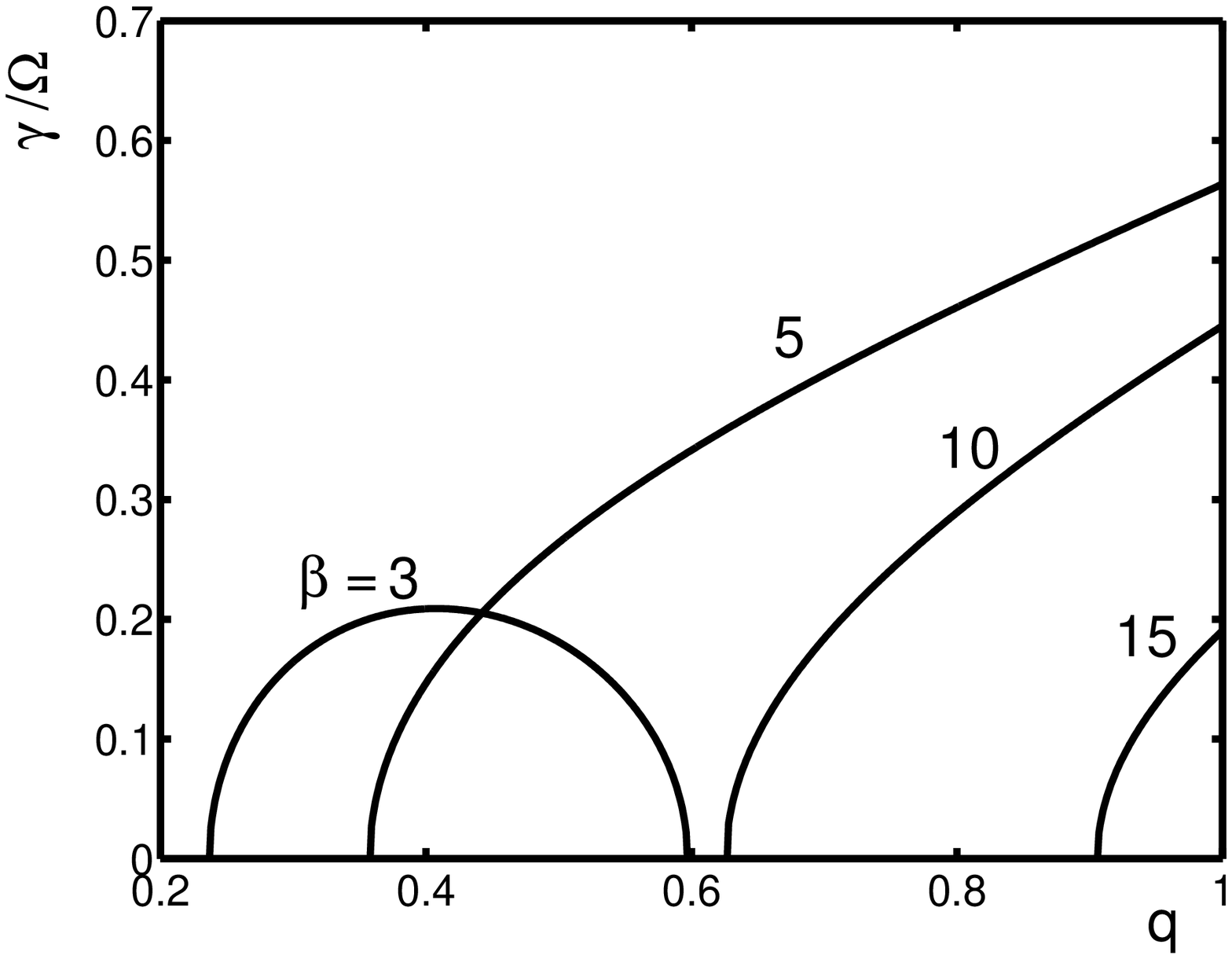} \caption{Growth rates of the Hall
instability for Keplerain rotation, and $\ell_i/L\sqrt {\zeta}
=10$ and $\xi =1.66$.
}
\end{figure}

\clearpage

\begin{figure}
\epsscale{.80} \plotone{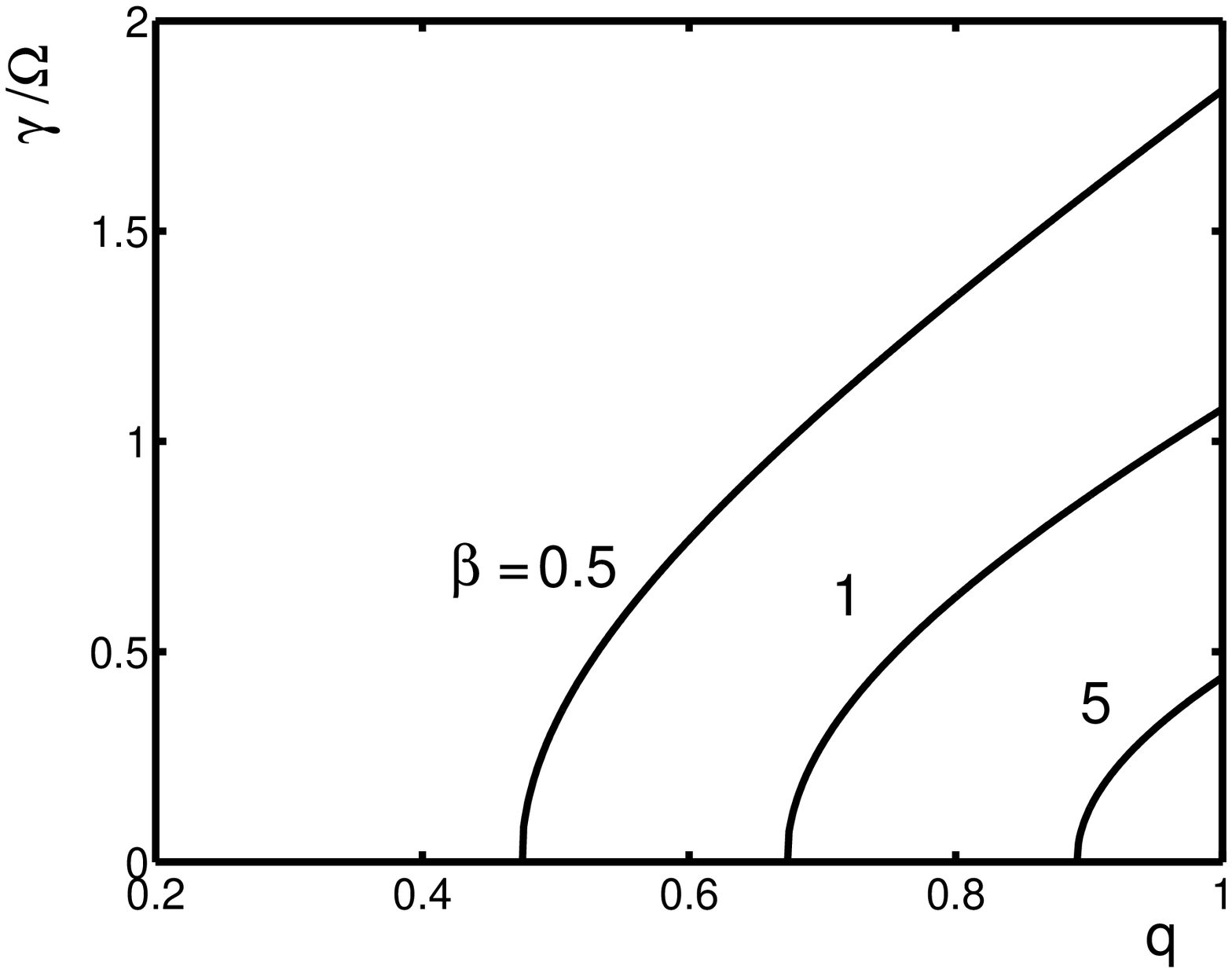} \caption{Growth rates of the Hall
instability for Keplerain rotation, and $\ell_i/L\sqrt {\zeta}
=10$ and $\xi =-1.5$.
\label{rootspic}}
\end{figure}







\end{document}